\documentclass[twocolumn,showpacs,showkeys,amsmath,amssymb,superscriptaddress,floatfix,nofootinbib]{revtex4}

\usepackage{graphicx}
\usepackage{bm} 
\usepackage{subfigure}

\def\vec#1{\mathchoice{\mbox{\boldmath$\displaystyle#1$}}
{\mbox{\boldmath$\textstyle#1$}}
{\mbox{\boldmath$\scriptstyle#1$}}
{\mbox{\boldmath$\scriptscriptstyle#1$}}}
\makeatletter
\newcommand\erfc{\mathop{\operator@font erfc}\nolimits}
\def\slashchar#1{\setbox0=\hbox{$#1$}
   \dimen0=\wd0 \setbox1=\hbox{/} \dimen1=\wd1
   \ifdim\dimen0>\dimen1 \rlap{\hbox to \dimen0{\hfil/\hfil}} #1
   \else  \rlap{\hbox to \dimen1{\hfil$#1$\hfil}} / \fi}

\begin{document}
 
\title{
Locally anisotropic momentum distributions of hadrons at freeze-out in relativistic heavy-ion collisions}

\author{Maciej Rybczy\'nski} 
\email{Maciej.Rybczynski@ujk.edu.pl}
\affiliation{Institute of Physics, Jan Kochanowski University, PL-25406~Kielce, Poland} 

\author{Wojciech Florkowski} 
\email{Wojciech.Florkowski@ifj.edu.pl}
\affiliation{Institute of Physics, Jan Kochanowski University, PL-25406~Kielce, Poland} 
\affiliation{The H. Niewodnicza\'nski Institute of Nuclear Physics, Polish Academy of Sciences, PL-31342 Krak\'ow, Poland}

\date{June 28, 2012}

\begin{abstract}
A spheroidal anisotropic local momentum distribution is implemented in the statistical model of hadron production. We show that this form leads to exactly the same ratios of hadronic abundances as the equilibrium distributions, if the temperature is identified with a characteristic transverse-momentum scale. Moreover, to a very good approximation the transverse-momentum spectra of hadrons are the same for isotropic and anisotropic systems, provided the size of the system at freeze-out is appropriately adjusted. We further show that this invariance may be used to improve the agreement between the model and experimental HBT results.
\end{abstract}

\pacs{25.75.-q, 25.75.Dw, 25.75.Ld}

\keywords{relativistic heavy-ion collisions, particle spectra, femtoscopy, LHC}

\maketitle 

\section{Introduction}
\label{sect:intro}

Thermal analyses of hadronic yields in heavy-ion collisions \cite{Koch:1985hk,Cleymans:1992zc,Csorgo:1995bi,Rafelski:1996hf,Rafelski:1997ab,Cleymans:1998fq,Cleymans:1999st,Gazdzicki:1998vd,Gazdzicki:1999ej,Braun-Munzinger:1994xr,Cleymans:1996cd,Becattini:2000jw,Sollfrank:1993wn,Schnedermann:1993ws,Braun-Munzinger:1995bp,Becattini:1997uf,Yen:1998pa,Braun-Munzinger:1999qy,Becattini:2003wp,Braun-Munzinger:2001ip,Florkowski:2001fp,Broniowski:2001we,Broniowski:2001uk,Broniowski:2002nf,Retiere:2003kf} have become a popular tool for the interpretation of the data collected at the AGS \cite{Braun-Munzinger:1994xr,Cleymans:1996cd,Becattini:2000jw}, SPS \cite{Sollfrank:1993wn,Schnedermann:1993ws,Braun-Munzinger:1995bp,Becattini:1997uf,Yen:1998pa,Braun-Munzinger:1999qy,Becattini:2003wp}, and RHIC energies \cite{Braun-Munzinger:2001ip,Florkowski:2001fp,Broniowski:2001we,Broniowski:2001uk,Broniowski:2002nf,Retiere:2003kf}. Most often, the successful results of such analyses are understood as the evidence for high-level thermalization of the hadronic systems produced in such collisions.

If the single-freeze-out scenario is assumed \cite{Broniowski:2001we,Broniowski:2001uk,Broniowski:2002nf}, in addition to the hadron yields, the thermal approach may be used to describe other physical observables. In particular, it has been successfully used to reproduce the RHIC transverse momentum spectra \cite{Florkowski:2002wn,Baran:2003nm}, collective flow \cite{Retiere:2003kf,Broniowski:2002wp,Florkowski:2004du}, femtoscopic observables \cite{Kisiel:2006is,Broniowski:2008vp,Kisiel:2008ws}, and charge balance functions \cite{Florkowski:2004em,Bozek:2003qi}. 

In the thermal approach, one usually obtains very good fits with just a few thermodynamic parameters, such as the temperature, $T$, the baryon and strangeness chemical potentials, $\mu_B$ and $\mu_S$, or in the extended approaches the quark fugacities, $\gamma_s$ and $\gamma_q$, and additional parameters describing the system's geometry and expansion (flow). 

Recently, the single-freeze-out model has been applied to the LHC data describing transverse-momentum spectra of hadrons produced in Pb+Pb collisions at the beam energy $\sqrt{s_{\rm NN}}=$~2.76 TeV \cite{Rybczynski:2012ed}. It has been shown that the model reproduces well the spectra of pions, kaons, and hyperons, however, it does not reproduce the proton spectra. This results agrees with an earlier finding that the ratios of all hadron abundances measured at the LHC, except for protons, may be described within the thermal model \cite{KalweitSQM}. 

In this paper we show that the agreement of the thermal models with the data does not necessarily imply the fact that the system is locally equilibrated. We take into account possible differences in the local distribution of {\it transverse} and {\it longitudinal} momenta (with the directions defined with respect to the beam axis). Using the covariant Romatschke-Strickland form (RSF) for the anisotropic phase-space momentum distributions, we show that i)~the ratios of hadron multiplicities are exactly the same as in the locally equilibrated system, if the temperature $T$ is identified with a characteristic transverse-momentum scale $\Lambda$, ii)~the transverse-momentum spectra of hadrons are changed, to a very good approximation, only by an overall factor which depends on the momentum anisotropy parameter appearing in RFS --- since this change may be easily compensated by a change of the parameters defining the size of the system, the transverse-momentum spectra may be also treated as insensitive to the discussed momentum anisotropy, finally, iii)~the discussed invariance of the spectra may be used to improve the model HBT results for pions. 

Generally speaking, our results demonstrate robustness of the thermal framework against specific variations of the model assumptions. This and other similar studies (see, for example, \cite{Michalec:2001qf}) show that the thermal approach is quite successful even in the situations where matter is out of equilibrium. Our results concerning the ratios of hadronic abundances agree qualitatively with the result of Ref. \cite{Schenke:2003mj}, where an approximate invariance of the ratios has been demonstrated with slightly different physical assumptions.

We stress that we consider the anisotropy between longitudinal and transverse momenta considered in the local rest frame of the fluid element. These two directions are defined always with respect to the beam axis. We do not consider the anisotropy of the transverse-momentum distributions in the $(p_x,p_y)$ plane, quantified by the harmonic flow coefficients $v_n$. 

For simplicity, in the discussion of the spectra we consider boost-invariant and cylindrically symmetric models that have been implemented in the Monte-Carlo generator {\tt THERMINATOR} \cite{Kisiel:2005hn,Chojnacki:2011hb}. We compare the results of the model calculations with the LHC data for Pb+Pb collisions at the beam energy $\sqrt{s_{\rm NN}}=2.76$~TeV.

\section{Cooper-Frye formalism and covariant Romatschke-Strickland form}
\label{sect:RS}

In this paper we assume that primordial hadrons are emitted from the freeze-out hypersurface $\Sigma$ and their spectra may be calculated with the help of the Cooper-Frye formula
\begin{equation}
E_p \frac{dN}{d^3p} =  \int d\Sigma_\mu(x) p^\mu f(x, p),
\label{cf}
\end{equation}
where $E_p$ is the particle's energy, $d\Sigma^\mu$ is the element of the freeze-out hypersurface, and $f(x, p)$ is the phase-space distribution function of the emitted particles. For systems in local thermal equilibrium one assumes Fermi-Dirac ($\epsilon=-1$) or Bose-Einstein ($\epsilon=+1$) statistics and uses the distribution functions
\begin{equation}
f_{\rm eq}(x, p) = g \left\{ \exp\left[\frac{p \cdot U -  \mu}{T}\right] -\epsilon \right\}^{-1}
\label{feq}
\end{equation}
where $g$ is the number of internal degrees of freedom, $T$ is the freeze-out temperature, and $\mu$ is the chemical potential. Typically, the freeze-out conditions are defined by the fixed values of $T$ and $\mu$, hence these two quantities may be treated as constants in (\ref{feq}).

The equilibrium distributions depend on the scalar product of the particle four-momentum $p^\mu$ and the four-velocity $U^\mu$ of the fluid element. We use the standard parameterizations
\begin{eqnarray}
\hspace{-0.5cm} p^\mu &=& \left(E_p, {\vec p}_\perp, p_\parallel \right) 
= \left(m_\perp \cosh y, {\vec p}_\perp, m_\perp \sinh y \right)  \label{pU}
\end{eqnarray} 
and
\begin{eqnarray}
U^\mu &=& \gamma (1, {\vec v}_\perp, v_\parallel), \label{pU}
\end{eqnarray} 
where $\gamma = (1-v^2)^{-1/2}$.

In order to study the effects of the anisotropic distributions, we modify Eq.~(\ref{feq}) and use \cite{Romatschke:2003ms,Florkowski:2011jg,Martinez:2012tu}
\begin{equation}
f_{\rm RSF}(x, p) = 
g \left\{ \exp\left[\frac{ \sqrt{(p \cdot U)^2 + \xi (p \cdot V)^2} }{\Lambda}\right] -\epsilon \right\}^{-1}.
\label{RSform}
\end{equation}
Here $\xi$ is the anisotropy parameter and $\Lambda$ is a typical transverse-momentum scale characterizing the particles in the system. For local thermodynamic equilibrium \mbox{$\xi=0$} and $\Lambda=T$ (from now on we set $\mu=0$).

The four-vector $V^\mu$ appearing in (\ref{RSform}) defines the direction of the beam and has the structure
\begin{equation}
V^\mu = \gamma_z (v_z, 0, 0, 1), \quad \gamma_z = (1-v_z^2)^{-1/2}.
\label{V}
\end{equation}
We note that the four-vectors $U^\mu$ and $V^\mu$ satisfy the normalization conditions
\begin{eqnarray}
U^2 = 1, \quad V^2 = -1, \quad U \cdot V = 0.
\label{UVnorm}
\end{eqnarray}
In the local rest frame (LRF) of the fluid element, $U^\mu$ and $V^\mu$ have simple forms
\begin{eqnarray}
 U^\mu = (1,0,0,0), \quad V^\mu = (0,0,0,1), 
 \label{UVLRF}
\end{eqnarray}
and the distribution function (\ref{RSform}) is reduced to
\begin{equation}
f_{\rm RSF}(x, p) = 
g \left\{ \exp\left[\frac{ \sqrt{m^2 + p_\perp^2 + (1+\xi) p_\parallel^2} }{\Lambda}\right] -\epsilon \right\}^{-1}.
\label{RSform1}
\end{equation}
Neglecting the hadron mass ($m=0$) and quantum statistics ($\epsilon=0$) we recover the commonly used RSF \cite{Romatschke:2003ms}.

\section{Ratios of hadronic yields}
\label{sect:ratios}

Starting with the Cooper-Frye formula (\ref{cf}) we obtain the multiplicity of the hadron species $i$ in the form 
\begin{equation}
N_i  =  \int d\Sigma_\mu(x)  \int \frac{d^3p}{E_p} p^\mu f_i(x, p).
\label{cfN}
\end{equation}
If the distribution function is taken as RSF, we have $f_i(x, p) = f^i_{\rm RFS}(p \cdot U,p \cdot V)$, and the integral over the momentum in (\ref{cfN}) yields the particle current
\begin{equation}
\int \frac{d^3p}{E_p} p^\mu f^i_{\rm RFS}(p \cdot U,p \cdot V) = n_i\left(\Lambda,\xi(x)\right) U^\mu.
\label{ni1}
\end{equation}
A few comments are in order now: i)~we assume now that the freeze-out is defined by the fixed value of the hard scale $\Lambda$, hence, $\Lambda$ is kept constant in (\ref{ni1}), ii)~we allow for the spacetime dependence of the anisotropy parameter $\xi$ but we assume that $\xi$ is the same for all hadronic species, iii)~there is no term proportional to $V^\mu$ on the right hand of Eq.~(\ref{ni1}) due to quadratic dependence of the distribution function on $p\cdot V$.

The calculation of the density $n_i\left(\Lambda,\xi(x)\right)$ may be done in the LRF and the result is
\begin{equation}
n_i\left(\Lambda,\xi(x)\right) = \frac{n_{i, \rm eq}\left(\Lambda\right)}{\sqrt{1+\xi(x)}}.
\label{ni2}
\end{equation}
where $n_{i, \rm eq}\left(\Lambda\right)$ is the particle density in equilibrium at the temperature defined by the parameter $\Lambda$.

Using Eq.~(\ref{ni2}) in (\ref{cfN}) we find
\begin{equation}
N_i = n_{i, \rm eq}\left(\Lambda\right) \int \frac{d\Sigma_\mu(x) U^\mu(x)}{\sqrt{1+\xi(x)}}
\end{equation}
and for the ratios
\begin{equation}
\frac{N_i}{N_j} = \frac{n_{i, \rm eq}\left(\Lambda\right) }{n_{j, \rm eq}\left(\Lambda\right) }.
\label{NiNj1}
\end{equation}

We thus see that the ratios of hadronic multiplicities are exactly the same as the ratios obtained for equilibrium distributions at the temperature $\Lambda$.

We also note that for the boost-invariant systems we have
\begin{equation}
\frac{N_i}{N_j} = \frac{\frac{dN_i}{dy} \Delta Y}{\frac{dN_j}{dy} \Delta Y}
= \frac{\frac{dN_i}{dy}}{\frac{dN_j}{dy}},
\end{equation}
where $\Delta Y$ is the rapidity range. Hence, in this case, the ratios of rapidity densities are the same as the ratios obtained for equilibrium at the temperature $\Lambda$.

\section{Transverse-momentum spectra}
\label{sect:spectra}

In this Section we calculate the transverse-momentum spectra in the thermal model with single freeze-out and compare our results with the LHC data. We use  {\tt THERMINATOR 2} \cite{Kisiel:2005hn,Chojnacki:2011hb} and choose two options for the shape of the freeze-out hypersurface: i) the Cracow model, and ii) the blast-wave model. These two models have been recently discussed in Ref. \cite{Rybczynski:2012ed}. In this work we follow closely the method presented in \cite{Rybczynski:2012ed}. The main difference is the use of the modified distributions at freeze-out.

\subsection{Cracow model}
\label{sect:cracow}

In the Cracow model the freeze-out hypersurface is defined by the condition that the invariant time \mbox{$(t^2-x^2-y^2-z^2)^{1/2}$} is fixed. The value of this time, $\tau_{\rm 3f}$, and the transverse size of the system, $r_{\rm max}$, are the two geometric parameters of the model. The flow of matter at freeze-out has the Hubble form $U^\mu = x^\mu/\tau_{\rm 3f}$.

If the hadronic system is in equilibrium, the freeze-out conditions are defined by the two extra parameters: the freeze-out temperature, $T$, and the freeze-out baryon chemical potential, $\mu$. In this work we neglect the chemical potential and introduce the scale parameter $\Lambda$ instead of $T$. In addition, the hadron distributions are taken as RSF with the anisotropy parameter $\xi$, which we keep constant in this Section. Thus, altogether we have four independent parameters: $\tau_{\rm 3f}, r_{\rm max}, \Lambda$ and $\xi$.

In Ref.~\cite{Rybczynski:2012ed} we analyzed Pb+Pb collisions at the beam energy $\sqrt{s_{\rm NN}}=2.76$~TeV for three centrality classes. We used three parameters: $\tau_{\rm 3f}, r_{\rm max}$ and $T$. The freeze-out temperature was always set equal to $T_0=165.6$~MeV, while the geometric parameters depended on the centrality class, we used: $\tau^0_{\rm 3f}=9.0$~fm and $r^0_{\rm max}=11.4$~fm for $c=$~0\%--5\%, $\tau^0_{\rm 3f}=7.4$~fm and $r^0_{\rm max}=9.6$~fm for $c=$~10\%--20\%, and finally $\tau^0_{\rm 3f}=5.9$~fm and $r^0_{\rm max}=7.25$~fm for $c=$~30\%--40\%.

\begin{figure}[t]
\begin{center}
\includegraphics[angle=0,width=0.5\textwidth]{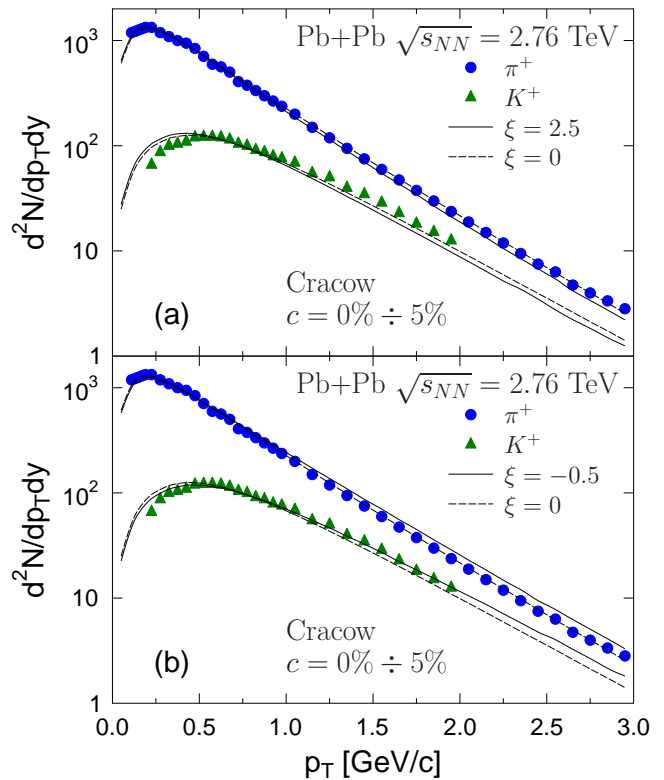}
\end{center}
\caption{\small (Color online) Transverse-momentum spectra of positive pions (dots) and kaons (triangles) measured in Pb+Pb collisions at the beam energy $\sqrt{s_{\rm NN}}=2.76$~TeV  \cite{Preghenella} compared to the Cracow model results with $\xi=2.5$ (a) and $\xi=-0.5$ (b). The anisotropic hadron distributions are used in the model calculations. The equilibrium parameters have been rescaled according to the formula (\ref{cracow-par}). The experimental and model results are shown for the centrality class $c=0\%-5$\%.}
\label{fig:fig_krakow}
\end{figure}

We have added the superscript $0$ to mark that the above values have been used in the equilibrium calculations. When switching from the equilibrium distributions  (\ref{feq}) to the anisotropic distributions (\ref{RSform}), we use the following scheme
\begin{eqnarray}
\Lambda &=& T_0 \nonumber \\
\tau_{\rm 3f} &=& \tau^0_{\rm 3f} (1+\xi)^{1/6}, \nonumber \\
r_{\rm max} &=& r^0_{\rm max} (1+\xi)^{1/6}.
\label{cracow-par}
\end{eqnarray}
The model transverse-momentum spectra obtained for this choice of the parameters are shown in Fig.~\ref{fig:fig_krakow} as solid lines and compared to the data. The part (a) describes the results for $\xi=2.5$, while the part (b) shows the results for $\xi=-0.5$. The results of the standard thermodynamic fit, obtained in Ref.~\cite{Rybczynski:2012ed}, are represented by the dashed lines. We clearly see that the change introduced by the momentum anisotropy, induced by the term $\xi (p\cdot V)^2$ in the exponential function, is very well compensated by the renormalization of the size of the system according to Eq.~(\ref{cracow-par}). 

We note that the power $1/6$ appearing in Eq.~(\ref{cracow-par}) implies that the volume of the system in the Cracow model scales as $(1+\xi)^{1/2}$, which compensates the same factor in the denominator of Eq.~(\ref{ni2}). Using the same rescaling for $\tau_{\rm 3f}$ and $r_{\rm max}$ we find that the transverse flow profiles change very moderately (within 20\%). Hence, the invariance of our results for the spectra with respect to the transformation (\ref{cracow-par}) is only approximate. This property is explained in more detail in the Appendix.

\subsection{Blast-wave model}
\label{sect:bwa}

In this Section we present our results obtained with the modified version of the blast wave model. This model differs from the Cracow model by the form of the freeze-out hypersurface and the form of the transverse flow. The shape of the freeze-out curve in the Minkowski space is controlled by the parameter $A$, whereas the magnitude of the flow is controlled by the parameter $v_T$. In the equilibrium version used in Ref.~\cite{Rybczynski:2012ed} we have used four parameters: $A$, $T_0$, $\tau^0_{\rm 2f} = r^0_{\rm max}$, and $v_T$. Their values are listed in Table~1 of~Ref.~\cite{Rybczynski:2012ed}.

\begin{figure}[t]
\begin{center}
\includegraphics[angle=0,width=0.5\textwidth]{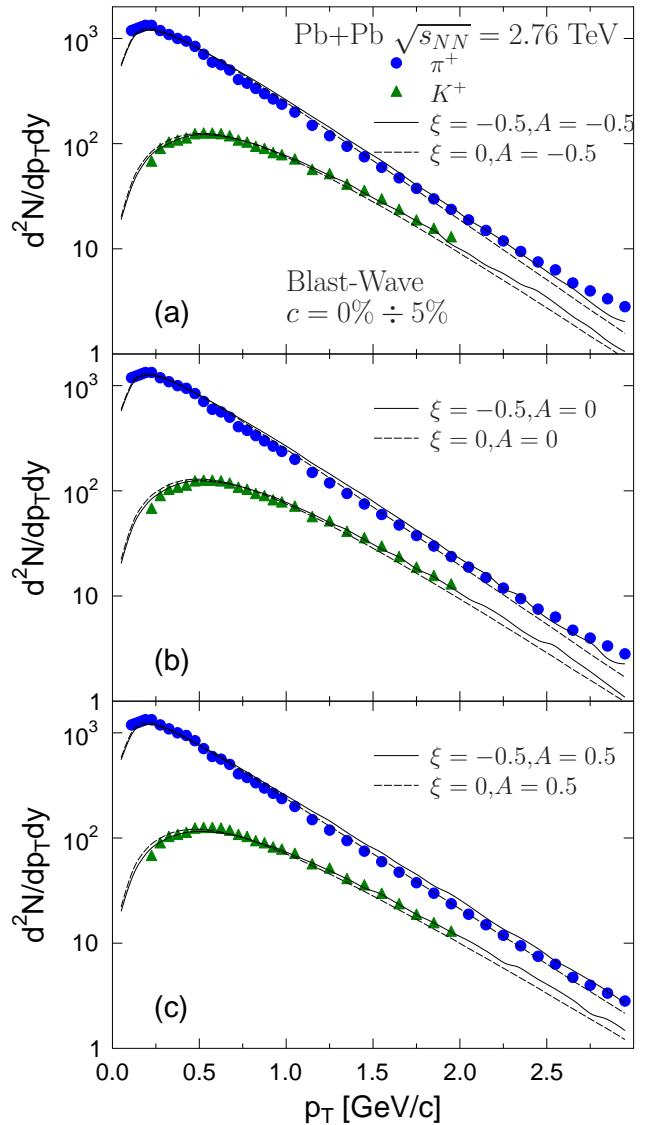}
\end{center}
\caption{\small (Color online) The transverse-momentum spectra obtained in the blast-wave model for three different values of the parameter $A$, $\xi=-0.5$ (solid lines) and $\xi=0$ (dashed lines). The equilibrium parameters have been rescaled according to the formula (\ref{bwa-par}). The data are taken from \cite{Preghenella}.
}
\label{fig:bwa_a}
\end{figure}

\begin{figure}[t]
\begin{center}
\includegraphics[angle=0,width=0.5\textwidth]{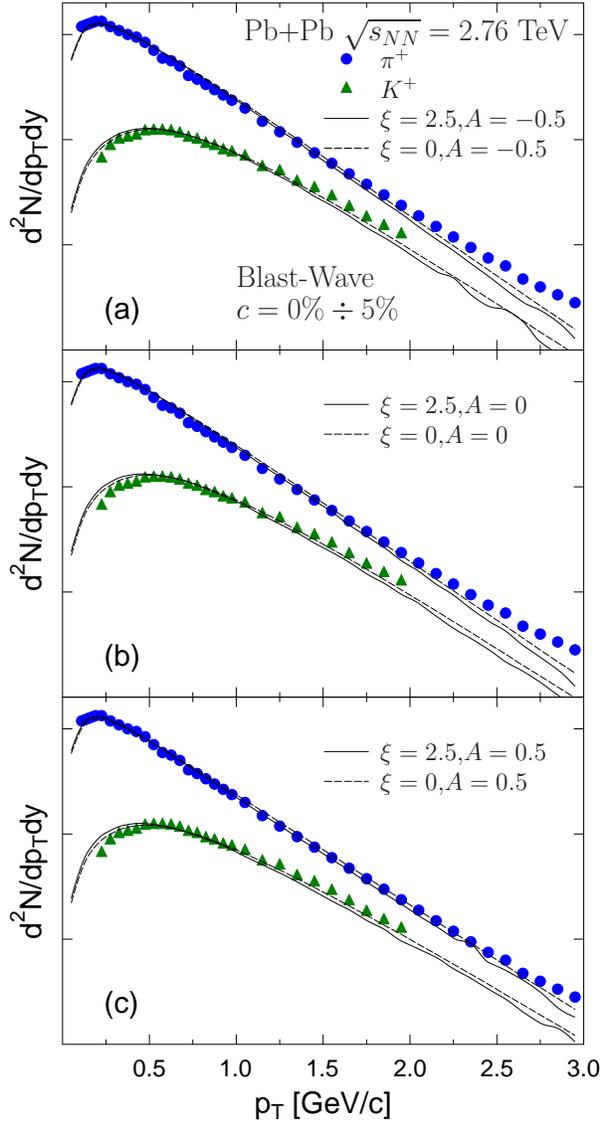}
\end{center}
\caption{\small (Color online) The same as Fig.~\ref{fig:bwa_a} but for $\xi=2.5$ (solid lines) and $\xi=0$ (dashed lines). The data are taken from \cite{Preghenella}.
}
\label{fig:bwa_b}
\end{figure}

Similarly to the case of the Cracow model, when switching from equilibrium to the anisotropic distributions, we make the following change of the parameters
\begin{eqnarray}
\Lambda &=& T_0 \nonumber \\
\tau_{\rm 2f} &=& \tau^0_{\rm 2f} (1+\xi)^{1/6}, \nonumber \\
r_{\rm max} &=& r^0_{\rm max} (1+\xi)^{1/6}.
\label{bwa-par}
\end{eqnarray}
Our results are shown in Figs.~\ref{fig:bwa_a}~and~\ref{fig:bwa_b}. Similarly to the Cracow model we observe that the spectra are (almost) insensitive to the induced anisotropy provided the geometric parameters of the model are appropriately rescaled.

\section{HBT radii}
\label{sect:bwa}

\begin{figure}[t]
\begin{center}
\includegraphics[angle=0,width=0.5\textwidth]{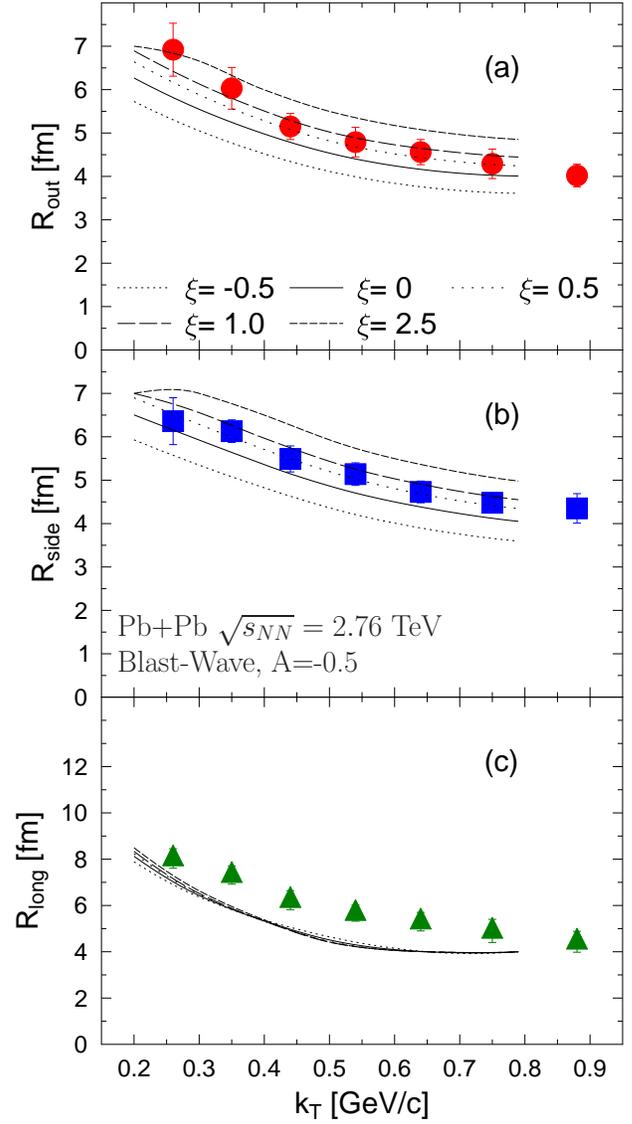}
\end{center}
\caption{\small (Color online) The pion HBT radii $R_{\rm out}$ (a), $R_{\rm side}$ (b), and $R_{\rm long}$  obtained in the blast-wave model with the freeze-out slope parameter $A=-0.5$ for five different values of the local momentum anisotropy $\xi$. The model results are compared to the ALICE data~\cite{Aamodt:2011mr}. }
\label{fig:HBT-neg-A}
\end{figure}

In Ref.~\cite{Rybczynski:2012ed} we have used the Cracow and the blast-wave model with different values of the parameter $A$ to calculate the pion HBT radii. The main outcome of these calculations is that the radii are reproduced best in the blast-wave model with $A=-0.5$. We recall that the freeze-out conditions described by a negative value of $A$ correspond to the situation where the outer parts of the system freeze out earlier than the system's interior. This is typical for more advanced hydrodynamic models.

In this Section we show the results of the calculations which are analogous to those presented in Ref.~\cite{Rybczynski:2012ed}. We choose the blast-wave model with $A=-0.5$ and incorporate the momentum anisotropy at freeze-out. For each value of the anisotropy parameter $\xi$ we modify the geometric parameters according to the formula (\ref{bwa-par}). As we have seen in the previous Section, the modification of the geometric parameters compensates the effect of the anisotropy in the transverse-momentum spectra. On the other hand, the change of the geometric parameters should affect the HBT radii, as they are connected with the space-time extensions of the system. 

Our results for the pion HBT radii obtained in the blast-wave model with $A=-0.5$ for several values of the anisotropy parameter $\xi$ are shown in Fig.~\ref{fig:HBT-neg-A} ($\xi=-0.5$ -- small dots, $\xi=0$ -- solid line, $\xi=0.5$ -- large dots, $\xi=1.0$ -- long dashes, $\xi=2.5$ -- short dashes). As expected, $R_{\rm side}$ grows with increasing values of $\xi$. This radius has a simple geometric interpretation of the transverse size of the system, hence, the increase of $R_{\rm side}$ reflects simply the growth of $r_{\rm max}$. We have found that the radius $R_{\rm out}$ behaves in the similar way. On the other hand, the radius $R_{\rm long}$ is practically independent of the anisotropy parameter $\xi$. The latter behavior may be understood as a net result of the two effects: a decrease of the longitudinal homogeneity length due to the decreased longitudinal pressure and an increase of this length due to the overall increase of the size of the system. Interestingly, the two effects almost completely cancel each other. 

The results represented by the solid lines correspond to the local equilibrium studied in Ref.~\cite{Rybczynski:2012ed}. They agree reasonably well with the HBT data. However, Fig.~\ref{fig:HBT-neg-A} shows that the agreement with the data may be improved if we introduce a non-zero anisotropy. A much better agreement with the data is obtained for $\xi=1.0$ than with $\xi=0$. 

The general conclusion from the HBT calculations presented in this Section is that the anisotropy of the momentum may be introduced as an additional characteristics of the system at freeze-out and its non-trivial value may be used to improve the agreement with the data. 

\section{Conclusions}
\label{sect:concl}

In this paper we have considered locally anisotropic momentum distributions of hadrons at freeze-out in relativistic heavy-ion collisions. We have taken into account the local anisotropy between the longitudinal and transverse momenta. Our results show that physical observables, such as the ratios of hadron abundances or the hadronic transverse-momentum spectra, are in practice indistinguishable from those obtained in an analogous equilibrium calculations --- the effect of the momentum anisotropy may be compensated by an appropriate change of the geometric models of the system. We have also demonstrated  that this freedom of the parameters may be used to improve the agreement of the model calculations with the measured HBT radii. 

Our results indicate insensitivity of the thermal approach against specific variations of the model assumptions and show that it may be quite successful even in the situations where matter is out of equilibrium. 

\begin{acknowledgements}

We thank M.~Strickland and R.~Ryblewski for discussions and critical comments concerning the manuscript. This work was supported by the Polish Ministry of Science and Higher Education under Grant No. N N202 263438, and National Science Centre, grant DEC-2011/01/D/ST2/00772.                                      

\end{acknowledgements}

\section{Appendix}
\label{sect:app1}

In this Section, we explain the approximate scaling of the transverse-momentum spectra. As an example, we consider the Cracow model. In the case of the blast-wave model, the arguments leading to the approximate scaling are similar. 

In the Cracow model, the spectra of primordial particles at zero rapidity are given by the integrals of the form
\begin{eqnarray}
&& {dN \over d{\tt y} d^2p_\perp} \label{hmdN1} \\
&& = { \tau_{\rm 3f \,}^3  \over (2\pi)^3}
\int\limits_0^{2\pi} d\phi 
\int\limits_{-\infty}^{\infty} d\eta
\int\limits_0^{\vartheta^{\rm max}_\perp} 
\,\, d\vartheta_\perp \hbox{cosh}\,\vartheta_\perp \hbox{sinh}\,\vartheta_\perp \nonumber \\
&& \times
\left[m_\perp \hbox{cosh}\vartheta_\perp \hbox{cosh}\eta  
- p_\perp \hbox{sinh}\vartheta_\perp \cos\phi \right] \nonumber \\
&& \times 
\exp\left[-\beta (m_\perp \hbox{cosh} \vartheta_\perp 
\hbox{cosh}\eta - p_\perp  \hbox{sinh}\vartheta_\perp \cos\phi) \right]. \nonumber 
\end{eqnarray}
Here $\beta=1/T$, $\phi$ is the azimuthal angle, $\eta$ is the space-time rapidity, and $\theta_\perp$ is the transverse rapidity connected with the transverse distance, $r = \tau_{\rm 3f \,} \sinh \theta_\perp$.

The momentum anisotropy is implemented by replacing $T$ by $\Lambda$ and by the change of the argument of the exponential function,
\begin{eqnarray}
&& m_\perp \hbox{cosh} \vartheta_\perp \hbox{cosh}\eta - p_\perp  
\hbox{sinh}\vartheta_\perp \cos\phi \to \nonumber \\
&& \sqrt{(m_\perp \hbox{cosh} \vartheta_\perp \hbox{cosh}\eta - p_\perp  
\hbox{sinh}\vartheta_\perp \cos\phi)^2 + \xi m_\perp^2 \sinh^2\eta}. \nonumber
\end{eqnarray}
The expression under the square root may be rewritten in the equivalent form as
\begin{eqnarray}
(p_\perp \cos\phi \cosh\theta_\perp -m_\perp \cosh\eta \sinh\theta_\perp)^2 \nonumber \\
+ m^2 + p^2_\perp \sin^2\phi + m_\perp^2 (1+\xi) \sinh^2\eta.
\end{eqnarray}
where $\cosh\eta$ is relatively suppressed by $\sinh\theta_\perp$. Since the main contribution to the integral (\ref{hmdN1}) comes from the region where $\eta \approx 0$, the hyperbolic functions may be replaced by their approximations, $\cosh\eta \approx 1$ and $\sinh\eta \approx \eta$. Then, one can easily see that the factor $1+\xi$ may be eliminated by the appropriate change of the integration variable, which leads to the overall change of the normalization of the spectra by the factor $(1+\xi)^{-1/2}$.


\end{document}